\documentclass[aip,jcp,reprint,amsmath,amssymb,floatfix,citeautoscript,nofootinbib]{revtex4-2}
\usepackage{cancel}
\usepackage{amsmath}
\usepackage{physics}
\usepackage{graphicx}
\usepackage{amssymb}
\usepackage{amsthm}
\usepackage{bm}
\usepackage{dcolumn}
\newcolumntype{x}[1]{D{.}{.}{#1}}

\usepackage{ragged2e}
\usepackage{txfonts}
\allowdisplaybreaks

\usepackage{color}
\usepackage[usenames,dvipsnames]{xcolor}
\definecolor{myblue}{rgb}{0,0,1}
\usepackage[breaklinks=true,colorlinks=true,linkcolor=myblue,urlcolor=myblue,citecolor=myblue]{hyperref}

\usepackage{tabularx}

\begin{document}

\title{
Performance of periodic EOM-CCSD for band gaps of inorganic semiconductors and insulators
}

\author{Ethan A. Vo}
\affiliation{Department of Chemistry, Columbia University, New York, NY 10027 USA}
\author{Xiao Wang}
\affiliation{Department of Chemistry and Biochemistry, University of California, Santa Cruz, CA 95064 USA}
\author{Timothy C. Berkelbach}
\email{t.berkelbach@columbia.edu}
\affiliation{Department of Chemistry, Columbia University, New York, NY 10027 USA}

\begin{abstract}
We calculate the band gaps of 12 inorganic semiconductors and insulators composed of atoms from the first three rows of the periodic table 
using periodic equation-of-motion coupled-cluster
theory with single and double excitations (EOM-CCSD). 
Our calculations are performed with atom-centered triple-zeta basis sets and up to 64 $k$-points
in the Brillouin zone.
We analyze the convergence behavior with respect to number of orbitals and number of $k$-points sampled, using
composite corrections and extrapolations to produce our final values.
When accounting for electron-phonon corrections to experimental band gaps, we find that EOM-CCSD has a mean signed error of
$-0.12$~eV and a mean absolute error of 0.42~eV; the largest outliers are C (error of $-0.93$~eV), BP ($-1.00$~eV), and
LiH ($+0.78$~eV).
Surprisingly, we find that the more affordable partitioned EOM-MP2 theory performs as well as EOM-CCSD.
\end{abstract}

\maketitle

\section{Introduction}

The accurate \textit{ab initio} prediction of electronic band structures and
band gaps is a major driving force behind the improvement and development of
increasingly accurate electronic structure methods. The failures of local and semilocal
density functional theory (DFT) for this task have long been
understood~\cite{Perdew1986,Seidl1996,MoriSanchez2008}, motivating the use of various hybrid
functionals~\cite{Muscat2001,Xiao2011,Garza2016}. In a many-body framework, the GW approximation to the
self-energy~\cite{Hedin1965,Hybertsen1985,Hybertsen1986} is arguably the most successful method on the basis of its good
accuracy and relatively low computational cost, with technical issues and extensions such as
self-consistency and vertex corrections continuing to be explored~\cite{Bruneval2006,vanSchilfgaarde2006,Shishkin2007a,Chen2015,Kutepov2022,Yeh2022}.

In the last decade, wavefunction-based electronic structure methods have been
increasingly applied to solids.
Seven years ago, McClain \textit{et al.}~\cite{McClain2017} performed the first
calculation of band gaps of periodic three-dimensional solids using
equation-of-motion coupled-cluster theory with single and double excitations
(EOM-CCSD) for ionization potentials and electron affinities, presenting
encouraging results for diamond and silicon. Since then, the method has been
applied to select solids, including MnO, NiO~\cite{Gao2020},
MoS$_2$~\cite{Pulkin2020}, and ZnO~\cite{Laughon2022}, and has
been developed as an impurity solver for dynamical mean-field
theory~\cite{Zhu2019,Shee2019,Zhu2019a}.  However, the statistical performance
of the method over a wide range of solids has yet to be established.
We note that EOM-CCSD has also been used to study the neutral excitation energies 
of bulk solids, in periodic~\cite{Lewis2020,Wang2020,Wang2021} and embedded cluster~\cite{Dittmer2019} frameworks,
and defects~\cite{Gallo2021,Lau2023}.

Here, we aim to conclude this first phase of exploratory work by presenting converged
band gaps of 12 simple, canonical semiconductors and insulators composed of atoms
from the first three rows of the periodic table.
Importantly, this work benefits from recent infrastructure developments,
including efficient calculation of periodic integrals and density fitting~\cite{Sun2017,Ye2021,Ye2021a}
and the development of Gaussian basis sets that can be reliably converged to
the basis set limit in closely packed solids~\cite{Ye2022}.
We hope that our results can help direct future research on the use
of wavefunction based methods for excitation energies of solids, such as
applications to more complex solids or improvements to cost or accuracy.

\section{Methods}

\subsection{Materials}

We study 12 semiconductors and insulators with a range of covalent and ionic bonding and diverse crystal structures,
including diamond (C, Si), zinc blende (SiC, BN, BP, AlP), rock salt (MgO, MgS, LiH, LiF, LiCL), and wurtzite
(AlN).
Each contains two atoms per unit cell except for AlN, which contains four atoms per unit cell.
We use the experimental lattice constants for all solids, which are given in Tab.~\ref{tab:gaps}.
For all solids, we use DFT with the PBE functional and a large $k$-point mesh to identify the points 
in the Brillouin zone where the valence band maximum and conduction band minimum occur.

\subsection{EOM-CCSD calculations}

Periodic EOM-CCSD calculations were performed using PySCF~\cite{Sun2017pyscf}
with Gaussian density fitting of two-electron
integrals~\cite{Sun2017,Ye2021,Ye2021a}. 
We perform a periodic Hartree-Fock calculation, then a ground-state CCSD calculation, and finally an
EOM-CCSD calculation for the ionization potential [IP, $E(N-1)-E(N)$] 
and the electron affinity [EA, $E(N+1)-E(N)$], where $E(N)$ is (formally) the ground state energy
of a system with $N$ electrons. The band gap
is the sum [IP+EA, $E(N+1)+E(N-1)-2E(N)$].
The total cost is dominated by the ground-state CCSD calculation, which 
has $O(N_k^4 o^2 v^4)$ scaling in CPU time and $O(N_k^3 o^2 v^2)$ scaling in storage, where
$N_k$ is the number of $k$-points sampled in the Brillouin zone and $o,v$ are the number of
occupied and virtual (unoccupied) orbitals per $k$-point.
These relatively high CPU and storage costs necessitate composite corrections and extrapolations
to make predictions in the complete basis set limit and thermodynamic limit (TDL).

\begin{figure*}
	\includegraphics[scale=0.85]{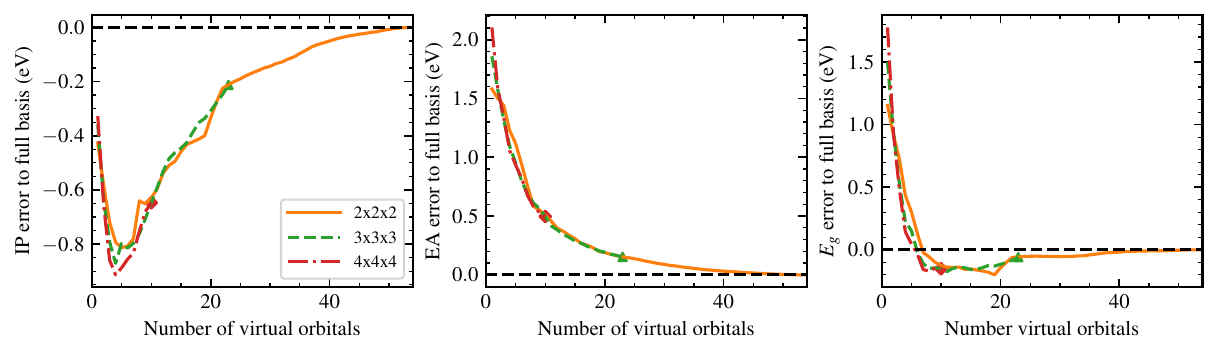}
	\caption{
Convergence of the IP, EA, and band gap with the number of correlated virtual orbitals
for silicon.  For $N_k=2^3$ data, the error is defined with respect to the full
TZ basis value (with  54  virtual orbitals per $k$-point).  The
$N_k=3^3$ and $4^3$ data have been shifted to align with that of the smaller
$k$-point mesh at the largest accessed number of virtual orbitals, which is 23 for
$3^3$ and 10 for $4^3$.
}
	\label{fig:basis}
\end{figure*}

We use GTH pseudopotentials optimized
for Hartree-Fock calculations~\cite{Goedecker1996,Hartwigsen1998} and the recently developed
correlation consistent GTH-cc-pVTZ basis set~\cite{Ye2021}.
Separate testing (not shown) with the QZ basis set confirm that basis set incompleteness errors due to our use of the TZ basis set are less than 0.1~eV in band gaps.
We sample the Brillouin zone using uniform $k$-point meshes with $N_k = 2^3$, $3^3$, and $4^3$.
For the larger $k$-point meshes, prohibitive storage costs require that we freeze virtual
orbitals [we always correlate all occupied (valence) orbitals, of which there are 4 per unit cell for all materials
studied here except LiH (1 occupied orbital) and AlN (8 occupied orbitals)]. 
We thus use a composite correction based on calculations with smaller $N_k$,
\begin{equation}
E(N_{k,2},\mathrm{L}) \approx E(N_{k,2},\mathrm{S}) + \left[ E(N_{k,1},\mathrm{L}) - E(N_{k,1},\mathrm{S})\right]
\end{equation}
where L and S indicate a large and small number of active virtual orbitals and $N_{k,2} > N_{k,1}$.
For $N_k=2^3$, we correlate the entire TZ basis set (except for AlN), 
which contains between 34 and 58 (67 for truncated AlN) orbitals per $k$-point; for $N_k=3^3$, we correlate 27 total orbitals
per $k$-point and correct based on calculations with $N_k=2^3$; for $N_k=4^3$, we correlate 14 total
orbitals per $k$-point and correct based on calculations with $N_k=3^3$.
Therefore, our calculations treat a few hundred electrons in about 1000 total orbitals, but benefit from the savings
implied by the translational symmetry of solids.

In Fig.~\ref{fig:basis}, we show the convergence of the (composite corrected) IP, EA, and band gap
as a function of the number of virtual orbitals, using silicon as an example. 
We see that the shape of convergence for all $N_k$ is
quite similar, suggesting that the composite corrections are accurate. The IP and EA have large, but opposite,
frozen orbital errors; these errors cancel in the band gap, which
converges to within 0.2~eV when correlating only 10 virtual orbitals per $k$-point and
to within 0.1~eV when correlating 23 virtual orbitals per $k$-point.
Beyond about 10 virtual orbitals, we see that the band gap converges from below, which is a general
trend seen in most other materials.

With these basis-set corrected estimates, we use a two-point extrapolation to the TDL
assuming finite-size errors that decay as $N_k^{-1/3}$,
\begin{equation}
E(N_k\rightarrow\infty) \approx \frac{N_{k,2}^{1/3}E(N_{k,2}) - N_{k,1}^{1/3}E(N_{k,1})}{N_{k,2}^{1/3}-N_{k,1}^{1/3}},
\end{equation}
where $N_{k,2} = 4^3$ and $N_{k,1} = 3^3$.
This slow convergence with $N_k$ is attributable to the finite-size error of a charged unit cell
and is shared by most many-body methods~\cite{Yang2020}.

\begin{figure}[b]
	\includegraphics[scale=0.9]{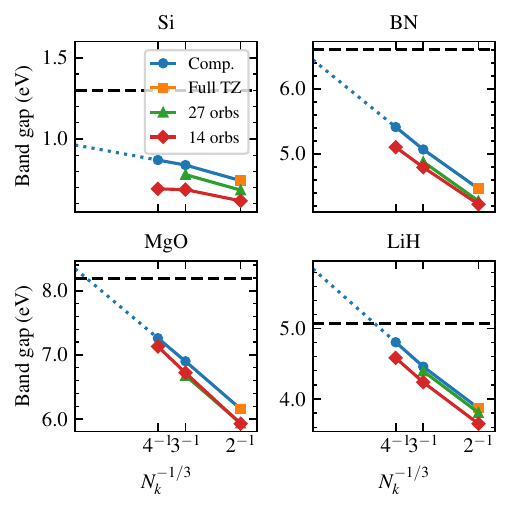}
	\caption{Behavior of the band gap as function of $N_k$ when correlating
the total number of orbitals indicated (14, 27, or the full TZ basis). The composite corrected curve (blue)
is our best estimate of the basis set limit, which is then extrapolated to the thermodynamic limit.
The black dashed line indicates the experimental band gap.
}
	\label{fig:gap_extrap}
\end{figure}

In Fig.~\ref{fig:gap_extrap}, we demonstrate this composite correction and extrapolation scheme for
four example solids: Si, BN, MgO, and LiH.
As already mentioned, we see that increasing the number of correlated orbitals increases the band gap.
We see that the difference in band gaps when correlating 27 or 14 orbitals is relatively independent
of $N_k$ (at least for $N_k=2^3$ and $3^3$), indicating that our additive composite correction
should be accurate.
For any given $N_k$, we estimate that our band gaps are basis-set converged to within about 0.1~eV,
although extrapolation can magnify these individual errors.
Moreover, the final basis-set corrected data (with $N_k = 2^3$, $3^3$, and $4^3$) do not always fall on a straight
line, suggesting that we may not have reached the limit where the finite-size error is exclusively due 
to the leading-order term of $O(N_k^{-1/3})$.

\subsection{P-EOM-MP2 calculations}

Because of the high cost of periodic EOM-CCSD calculations, it is worthwhile to compare to more affordable
theories such as MP2, i.e., the calculation of IPs, EAs, and band gaps using the second-order self-energy
with a Hartree-Fock reference.
However, MP2 has been shown to perform poorly for band gaps, predicting negative band gaps for semiconductors
like silicon~\cite{Marsman2009,Grueneis2010,Lange2021}.
Instead, our group recently explored
the accuracy of the closely related partitioned EOM-MP2 (P-EOM-MP2) theory~\cite{Lange2021}.
In P-EOM-MP2, we make two approximations to an EOM-CCSD calculation. First, we replace the ground-state
CCSD amplitudes by their second-order approximations. Second, we replace the large
doubles-doubles block of the similarity transformed Hamiltonian by a diagonal matrix, keeping only
the orbital energy differences.
These approximations lower the CPU and memory costs significantly.

A diagrammatic analysis in Ref.~\onlinecite{Lange2021} showed that a P-EOM-MP2 calculation 
essentially corresponds to supplementing the second-order self-energy with two third-order diagrams.
Remarkably, this minor difference resulted in significantly improved performance.
In this work, we will compare our EOM-CCSD results to P-EOM-MP2 results, which have
been recalculated here for consistency with the pseudopotentials, basis sets,
and $k$-points used in this work. However, we find numbers in good agreement
with our previous work~\cite{Lange2021}, suggesting that errors due to these
latter technical details are small.

\section{Results}

\subsection{Comparing EOM-CCSD and P-EOM-MP2}

\begin{figure}[t]
	\includegraphics[scale=0.9]{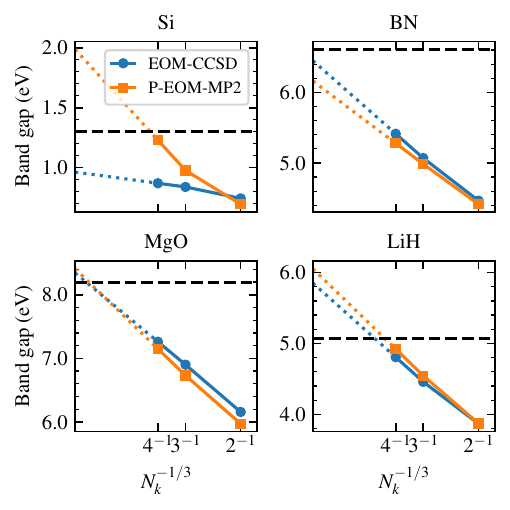}
	\caption{Comparison of EOM-CCSD and P-EOM-MP2 band gaps upon extrapolation to the thermodynamic limit.
        The black dashed line indicates the experimental band gap.}
	\label{fig:ccsd_mp2_nk}
\end{figure}

\begin{figure*}[t]
	\includegraphics[scale=0.85]{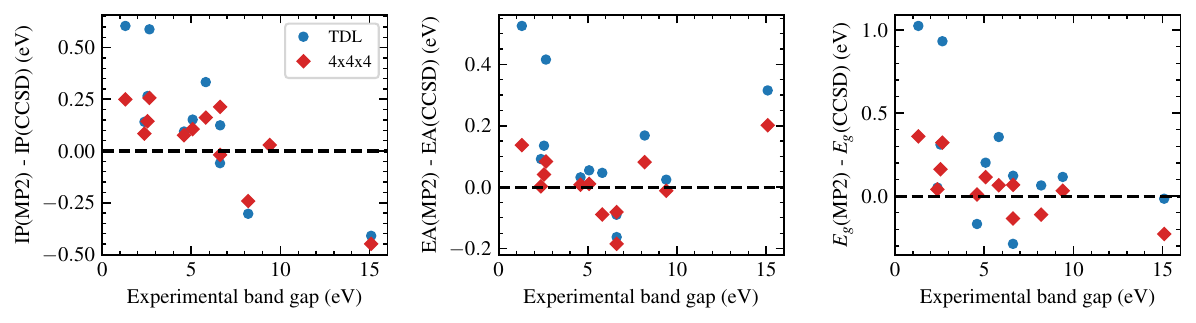}
	\caption{Difference between P-EOM-MP2 and EOM-CCSD for the IP, EA, and band gap $E_\mathrm{g}$ of all materials studied
        in this work. Differences are shown for values calculated with a $4\times 4\times 4$ $k$-point mesh 
        and for values extrapolated to the thermodynamic limit (TDL).}
	\label{fig:ccsd_mp2_diff}
\end{figure*}

Before comparing to experimental band gaps, we first compare the performance of EOM-CCSD and P-EOM-MP2. 
In Fig.~\ref{fig:ccsd_mp2_nk}, we compare the (composite corrected) band gaps as a function of $N_k$
for four example materials. Clearly, for many materials, the calculated band gaps are very similar:
at fixed $N_k$, the gaps commonly differ by less than 0.2~eV, which is a significant finding given the
difference in cost between the two methods.

However, for silicon (and a few other materials, see below), we see large differences that are magnified on approaching
the TDL. With $N_k = 4^3$, the EOM-CCSD gap is smaller by 0.4 eV, and extrapolation to the TDL
magnifies this difference to 1.0~eV. We note that our EOM-CCSD band gap in the TDL
is 0.96~eV, which is slightly smaller than the previously reported value of 
1.19~eV~\cite{McClain2017}.
We have confirmed that this discrepancy is due to a combination of small errors in our composite correction and differences in our basis set, pseudopotential, and $k$-point used for the conduction band minimum.

We next sought to identify any trend in the difference between P-EOM-MP2 and EOM-CCSD band gaps.
In Fig.~\ref{fig:ccsd_mp2_diff}, we show the difference of the IP, EA, and band gap as a function
of the experimental band gap; because extrapolation sometimes alters the difference, we show differences
at $N_k = 4^3$ and in the extrapolated TDL. We identify the following rough trends. The P-EOM-MP2 IP is larger
for small gap materials and smaller for large gap materials; the P-EOM-MP2 EA is larger for most materials;
the P-EOM-MP2 band gap is larger for most materials (typically by less than 0.5~eV), but similar for large gap materials.
The similarity of the band gaps for large materials is consistent with their more weakly correlated nature.
Overall, we find that P-EOM-MP2 and EOM-CCSD predict surprisingly similar band gaps. The largest outliers occur
for small gap materials upon extrapolation, and these are Si and BP, for which P-EOM-MP2 predicts
a band gap that is larger by about 1~eV.

\subsection{Comparing to experiment}

\begin{table*}[t]
	\centering
	\begin{tabular*}{0.98\textwidth}{@{\extracolsep{\fill}} lcdddddd}
\hline\hline
& & \multicolumn{2}{c}{Reference} & \multicolumn{2}{c}{P-EOM-MP2} & \multicolumn{2}{c}{EOM-CCSD} \\
Material    & \multicolumn{1}{c}{$a$ ($c$) (\AA)}	& \multicolumn{1}{c}{Expt. $E_\mathrm{g}$ (corrected)} & \multicolumn{1}{c}{el-ph} & \multicolumn{1}{c}{$E_\mathrm{g}$} & \multicolumn{1}{c}{Error} & \multicolumn{1}{c}{$E_\mathrm{g}$} & \multicolumn{1}{c}{Error} \\
		\hline
  Si & 5.431        & 1.30 & -0.06          &  1.99 &  0.69 &  0.96 & -0.34 \\
 SiC & 4.350        & 2.37 & -0.17          &  2.59 &  0.22 &  2.54 &  0.17 \\
 AlP & 5.451        & 2.54\cite{Lorenz1970} & -0.09          &  2.93 &  0.39 &  2.62 &  0.08 \\
  BP & 4.538        & 2.65 & -0.25\cite{Ha2020} &  2.58 & -0.07 &  1.65 & -1.00 \\
 MgS & 5.191        & 4.59\cite{Ching1995} &  -          &  5.10 &  (0.51) &  5.26 &  (0.67) \\
 LiH & 4.083        & 5.07\cite{Baroni1985} &  -0.08\cite{Monserrat2013} &  6.05 &  0.98 &  5.85 &  0.78 \\
  C  & 3.567        & 5.81 & -0.33          &  5.24 & -0.57 &  4.88 & -0.93 \\
  BN & 3.615        & 6.61 & -0.41          &  6.16 & -0.45 &  6.45 & -0.16 \\
 AlN & 3.110~(4.980)& 6.62 & -0.42         &  6.45 & -0.17 &  6.33 & -0.29 \\
 MgO & 4.213        & 8.19 & -0.52          &  8.41 &  0.22 &  8.34 &  0.15 \\
LiCl & 5.130        & 9.40\cite{Brown1970} &  -          &  9.55 &  (0.15) &  9.43 &  (0.03) \\
 LiF & 4.035        & 15.09 & -0.59         & 15.41 &  0.32 & 15.43 &  0.34 \\
\hline
MSE (eV) &              &      &                &       & 0.16 &       & -0.12 \\
MAE (eV) &              &      &                &       & 0.41 &       &  0.42 \\

		\hline\hline
\end{tabular*}
	\caption{Lattice constants and electronic band gaps of the 12 materials studied in this work (all energies in eV). Experimental band gaps have been corrected for the effects of electron-phonon coupling, when such corrections are available in literaure.
Except where indicated, experimental band gaps have been taken from the collection in Ref.~\onlinecite{Zhu2021}, 
and electron-phonon corrections have been taken from the collection in Ref.~\onlinecite{Lange2021}.
} 
	\label{tab:gaps}
	\end{table*}

We now turn to a comparison between calculated and experimental band gaps.
Our basis-set corrected and TDL extrapolated results for the electronic band
gap are given in Tab.~\ref{tab:gaps}, where they are compared to experimental band
gaps. For a fair comparison, we have corrected the experimental band gaps for
finite-temperature and vibrational zero-point energy effects, which typically act
to reduce the purely electronic band gap. 
If we exclude solids without available electron-phonon corrections, EOM-CCSD predicts band gaps with
a mean signed error (MSE) of $-0.12$~eV and a mean absolute error (MAE) of 0.42~eV. 
The largest outliers are diamond (error of $-0.93$~eV), BP ($-1.00$~eV), and LiH ($+0.78$~eV).
The statistical performance of P-EOM-MP2 is quite similar, with an MSE of 0.16~eV and MAE of 0.41~eV.

\begin{figure}[t]
	\includegraphics[scale=0.9]{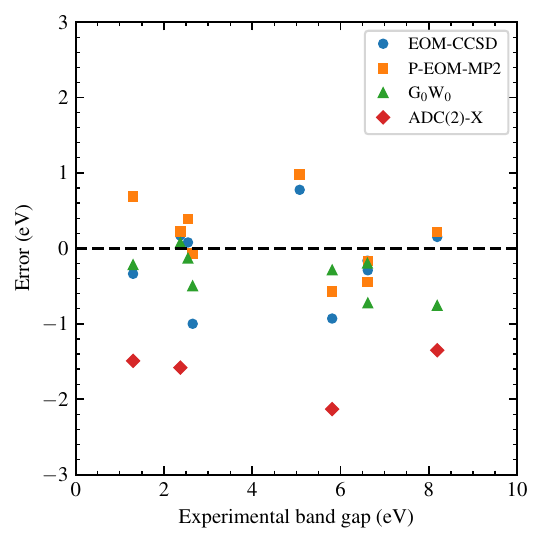}
	\caption{Band gap error for the 12 semiconductors and insulators studied in this work. In addition to our
own EOM-CCSD and P-EOM-MP2 results, we include ADC(2)-X results~\cite{Banerjee2022} and G$_0$W$_0$ results~\cite{Zhu2021}
from previous works, for comparison; only some of the same solids were studied in these previous works.}
	\label{fig:error_exp}
\end{figure}

The performance of both methods is shown graphically in Fig.~\ref{fig:error_exp}, along with that
of two other Green's function based methods, whose band gaps were previously published: 
the extended second-order algebraic diagrammatic construction [ADC(2)-X]~\cite{Banerjee2022} 
and the G$_0$W$_0$ approximation with a PBE reference~\cite{Zhu2021}.
We choose these two for comparison because their calculations were also performed with PySCF, using
similar basis sets and pseudopotentials.
We see that EOM-CCSD and P-EOM-MP2 outperform ADC(2)-X, which predicts band gaps that are too small, but
are comparable to the G$_0$W$_0$ approximation.

\section{Conclusions}

Through a study of 12 simple semiconductors and insulators, we have found that EOM-CCSD predicts band gaps
with a mean absolute error of about 0.4~eV. Perhaps unsurprisingly, this accuracy is very similar to our
group's previous finding concerning the accuracy of EOM-CCSD for neutral excitation energies of solids~\cite{Wang2020,Wang2021}.
Overall, we conclude that the performance of EOM-CCSD is not measurably better than that of P-EOM-MP2, even for small gap
semiconductors that one might expect to be more strongly correlated.

Although we believe our results are converged (with respect to basis set and number of $k$-points) to about 0.1--0.2~eV,
the need for composite corrections and extrapolation introduces uncertainties and prevents routine use of these correlated methods.
More robust methods for the reduction of basis set errors and finite-size errors would be very valuable.
Having established the performance of band gaps of simple solids, future work should explore more complex solids as well as core ionization
energies and the charged excitations of metals. The degree to which EOM-CCSD can replace or complement existing and more affordable
methods remains to be seen. Predicting very precise excitation energies through the incorporation of full or perturbative triples---a manner
of systematic improvability that is not shared by most existing methods---is perhaps the most exciting near-term goal.
Addressing the cost is the outstanding challenge.

\end{document}